\documentclass[iop]{emulateapj}

\newcommand{\um}{~\mu m}
\newcommand{\msun}{M_{\odot}}
\newcommand{\Lp}{L^\prime}

\usepackage{url}
\usepackage{xcolor}
\usepackage{hyperref}
\hypersetup{
    colorlinks = true,
    citecolor = blue,
    linkcolor = blue,
}

\begin{document}

\shorttitle{A Planetary-Mass Companion to HD~106906}
\shortauthors{Bailey et al.}

\title{HD~106906~\lowercase{b}: A planetary-mass companion outside a massive debris disk}

\author{Vanessa Bailey\altaffilmark{1}, 
Tiffany Meshkat\altaffilmark{2},
Megan Reiter\altaffilmark{1},
Katie Morzinski\altaffilmark{1}\footnotemark[*],
Jared Males\altaffilmark{1}\footnotemark[*],
Kate Y.\ L.\ Su\altaffilmark{1}, 
Philip M.\ Hinz\altaffilmark{1},  
Matthew Kenworthy\altaffilmark{2},
Daniel Stark\altaffilmark{1},
Eric Mamajek\altaffilmark{3},
Runa Briguglio\altaffilmark{4},
Laird M.\ Close\altaffilmark{1},
Katherine B.\ Follette\altaffilmark{1},
Alfio Puglisi\altaffilmark{4},
Timothy Rodigas\altaffilmark{1,5},
Alycia J.\ Weinberger\altaffilmark{5},
and
Marco Xompero\altaffilmark{4}
}

\affil{$^1$ Steward Observatory, University of Arizona, 933 North Cherry
Avenue, Tucson, AZ 85721, USA; {vbailey@as.arizona.edu}}
\affil{$^2$ Leiden Observatory, Leiden University, P.O. Box 9513, 2300 RA Leiden, The Netherlands}
\affil{$^3$ Department of Physics and Astronomy, University of
Rochester, Rochester, NY 14627-0171, USA}
\affil{$^4$ Osservatorio Astrofisico di Arcetri, Largo Enrico Fermi 5, I-50125 Firenze, Italy}
\affil{$^5$ Carnegie Institution of Washington, Department of Terrestrial Magnetism, 5241 Broad Branch Road NW, Washington, DC 20015, USA}

\footnotetext[*]{NASA Sagan Fellow}

\slugcomment{Accepted to ApJL}

\begin{abstract}
We report the discovery of a planetary-mass companion, HD~106906~b, with the new Magellan Adaptive Optics (MagAO) + Clio2 system. The companion is detected with Clio2 in three bands: $J$, $K_S$, and $\Lp$, and lies at a projected separation of 7\farcs1 (650~AU). It is confirmed to be comoving with its $13\pm2$~Myr-old F5 host using \textit{Hubble Space Telescope}/Advanced Camera for Surveys astrometry over a time baseline of 8.3~yr. DUSTY and COND evolutionary models predict the companion's luminosity corresponds to a mass of $11\pm2~M_{Jup}$, making it one of the most widely separated planetary-mass companions known.  We classify its Magellan/Folded-Port InfraRed Echellette $J/H/K$ spectrum as L$2.5\pm1$; the triangular $H$-band morphology suggests an intermediate surface gravity. HD~106906~A, a pre-main-sequence Lower Centaurus Crux member, was initially targeted because it hosts a massive debris disk detected via infrared excess emission in unresolved \textit{Spitzer} imaging and spectroscopy. The disk emission is best fit by a single component at 95~K, corresponding to an inner edge of 15--20~AU and an outer edge of up to 120~AU. If the companion is on an eccentric ($e>0.65$) orbit, it could be interacting with the outer edge of the disk. Close-in, planet-like formation followed by scattering to the current location would likely disrupt the disk and is disfavored. Furthermore, we find no additional companions, though we could detect similar-mass objects at projected separations $>35$~AU. In situ formation in a binary-star-like process is more probable, although the companion-to-primary mass ratio, at $<1\%$, is unusually small.

\end{abstract}

\keywords{instrumentation: adaptive optics --- open clusters and associations: individual (Lower Centaurus Crux) --- planet-disk interactions --- planetary systems --- stars: individual (HD 106906)}

\section{Introduction}

The handful of known planetary-mass companions at tens to hundreds of AU are already challenging planet formation theories, thus each addition to the set of directly imaged (DI) companions is valuable for understanding formation mechanisms. DI surveys are resource-intensive, as fewer than $20\%$ of stars have giant planets at large orbital separations \citep[e.g.][]{Vigan2012, Nielsen2013}. Therefore, there is a strong incentive to find so-called ``signposts'' for planets. 

Systems like HR~8799 and $\beta$~Pic host both planets and debris disks, with the planets likely sculpting the disks \citep{Su2009, Lagrange2010}. Several DI surveys \citep[e.g.][]{Apai2008} have targeted debris-disk-hosting stars and found planet occurrence rates comparable to disk-blind surveys. However, these groups did not have or did not utilize detailed information on the debris disk morphology.  We, and others \citep[e.g.][]{Janson2013a, Wahhaj2013}, hope to improve the odds by searching for planets in systems with unusual debris disks. We are targeting systems with infrared (IR) spectral energy distributions (SEDs) indicative of disk configurations such as two-belt \citep{Su2013} and large inner cavity systems. HD~106906 falls into the second category. 

HD~106906 (HIP~59960) is a member of the Lower Centaurus Crux (LCC) association, based on \textit{Hipparcos} kinematics \citep{deZeeuw1999}. The cluster has a mean age of $17$~Myr, with an age-spread of $\sim10$~Myr. HD~106906 is a negligibly reddened, pre-main-sequence F5V-type star, with an isochronal age and mass of $13\pm2$~Myr and $1.5~\msun$ \citep{Pecaut2012}.

In this Letter we present the first discovery of a planetary-mass companion around a debris-disk-selected star with the Magellan Adaptive Optics (MagAO) + Clio2 system. In Section \ref{sect:observe} we describe our observations with the Clio2 and FIRE instruments. In Section \ref{sect:A&R}, we: confirm common proper motion using Gemini NICI and \textit{Hubble Space Telescope} archival data; present near-infrared (NIR) spectroscopy of the companion to confirm its cool, young nature; estimate its mass using ``hot start'' evolutionary models; place limits on the presence of additional objects; and discuss the likelihood of interaction with the debris disk surrounding the primary star.

\section{Observations and Data Reduction}
\label{sect:observe}

\begin{deluxetable*}{llccccll}
\tablecaption{Summary of Observations.\label{tab:obs}}
\tablewidth{0pt}
\tablehead{
\colhead{Date}  & \colhead{Inst.\ }  & \colhead{Mode} & \colhead{Rot.\ $(\degr)$} & \colhead{Filter}  & \colhead{Bandpass ($\mu$m)} & \colhead{Exposure (sec)} & \colhead{Total Int.\ (min)}   
}  
\startdata
2013 Apr 4  &  Clio2 & ADI & 62 & $\Lp$  &  $3.41-4.10$ & 0.800/0.164\tablenotemark{a} & 80 (19\tablenotemark{b}) \\
2013 Apr 12 &  Clio2 & ADI & 9  & $J$     &  $1.17-1.33$ & 30/0.164\tablenotemark{a} & 9 (4.5\tablenotemark{b})  \\
2013 Apr 12 &  Clio2 & ADI & 9  & $K_S$   &  $2.00-2.30$ & 0.280 &  7  \\
2013 May 1 &  FIRE  & Track & - &  $J-K_S$  &  $1.0-2.5$   & 120   & 8 (4\tablenotemark{b})  \\
\\
2004 Dec 1 &  ACS    & Track & - &  $F606W$  & $0.47-0.71$  & 2500  & 42  \\
2011 Mar 21 &  NICI   & Track & - &  $H_2$    & $2.11-2.14$  & 80.2/0.38\tablenotemark{a}  & 1.3  \\
2011 Mar 21 &  NICI   & Track & - &  $K_S$    & $2.00-2.30$  & 80.2/0.38\tablenotemark{a}  & 1.3  \\
\enddata

\tablenotetext{a}{Exposure times for saturated/unsaturated images.} 
\tablenotetext{b}{Because of nodding, field rotation, and/or optical ghosts, ``b'' was only visible and/or detected at high S/N in this subset of the data.}

\end{deluxetable*}

\subsection{Clio2}

We used the Clio2 $1-5\um$ camera \citep{Sivanandam2006} behind the new MagAO natural guide star AO system \citep{Close2013} on the 6.5~m Magellan Clay telescope. MagAO/Clio2 is optimized for thermal IR wavelengths ($3-5\um$), where star-to-planet contrast is minimized \citep{Burrows1997b}. Clio2 has a plate scale of $15.86\pm0.05$~mas~px$^{-1}$ and a field of view (FOV) of $5''\times16''$ in the magnification and subarray mode selected, based on 2013 April 7 astrometric observations of the central stars of the Trapezium \citep{Close2012}. There may be systematic errors from distortion of up to $0.4\%$ in plate scale and $0\fdg2$ in rotation (based on Trapezium data). Data were obtained on 2013 April 4 and 12; conditions on both nights were photometric with winds of $7-11$~m~s$^{-1}$ and seeing of 1\arcsec or less. Observations were taken in Angular Differential Imaging mode \citep[ADI;][]{Marois2006}, with a two-position nod plus dither pattern ($3-6''$ nods, $\sim1''$ dithers). At each nod position, long science exposures and a short calibration frame were obtained. Further details are listed in Table \ref{tab:obs}. In the $L^\prime$ data, the companion was only within the FOV in one of the two nod positions. In the $J$ data, the companion was contaminated by an optical ghost in one of the nod positions; these data were discarded.

The data were processed with a custom reduction script\footnote{For image alignment, the offset for each image was calculated independently.} and aperture photometry was used, following \citet{Bailey2013}. The resulting images are shown in Figure \ref{fig:Imgs}. At separations of $0\farcs3-3''$, we fit and removed the contribution from the stellar point-spread function (PSF) using principal component analysis (PCA) with 15 principal components, as described in \citet{Meshkat2013a}. To quantify the effect of PCA analysis on point sources, artificial planets were injected into the raw data. The signal-to-noise ratio (S/N) of the recovered sources was determined following \citet{Bailey2013} and was used to infer the $5\sigma$ contrast as a function of separation. Our two-position nod observing strategy resulted in asymmetric sky coverage and thus decreasing sensitivity at increasing separations.

\subsection{FIRE}

We obtained an NIR spectrum of HD~106906~b on 2013 May 1 with the Folded-Port InfraRed Echellette \citep[FIRE;][]{Simcoe2013} on the Baade 6.5~m Magellan telescope (Table \ref{tab:obs}). FIRE simultaneously captures 0.8 - 2.5 \micron\ spectra with moderate resolution ($R=4800$). The FIRE echelle slit is 7\arcsec\ long; we chose a slit width of $0\farcs75$ to accommodate the $\sim0\farcs8$ seeing. Weather conditions otherwise were poor with high winds and thin clouds. The data were reduced using the FIRE reduction pipeline. A faint background contaminant, probably stellar, fell in the slit at a projected distance of $\sim2\farcs5$, and was masked during processing. HD~106906~b was below the detection limit at $\lambda<1.25\um$ in this short observation. Increasing sky and instrumental backgrounds decreased the data quality at $\lambda>2.25\um$. The resulting spectrum (binned to $R\sim500$) is presented in Figure \ref{fig:spectrum}; the S/N varies between three and six and is highest in the $H$-band.

\subsection{Ancillary Observations}

To confirm the comoving status and red color of the candidate, we used archival data from \textit{Hubble Space Telescope} Advanced Camera for Surveys (\textit{HST}/ACS). ACS took coronagraphic observations on 2004 December 1 (program 10330, PI: H.\ Ford; see Table \ref{tab:obs}). We used World Coordinate System information encoded in the standard distortion-corrected ACS archive data products. The primary's location beneath the coronagraphic mask was determined from its position relative to several background stars also seen in a non-coronagraphic acquisition frame; uncertainties in this measurement dominate the astrometric error budget. For photometric measurements of the companion we analyzed the standard cleaned, flat-fielded ACS archive data products with the \textit{Starfinder} PSF fitting routine and TinyTim synthetic PSF.

We also utilized Gemini/NICI data from 2011 March 21 (program GS-2011A-Q-44, PI: R.\ Jayawardhana; see Table \ref{tab:obs}). In this $K_S$ and $H_2$-band snapshot program of young stars in the Sco-Cen association \citep{Janson2013b}, two images were obtained per target (long and short exposures). Multiple targets were observed consecutively, with the stars placed at different locations on the detector; we performed sky subtraction using exposures from the target observed immediately prior to HD~106906. Images were flat-fielded using calibration data from 2011 March 30 and distortion-corrected using the polynomials provided by the observatory. The primary was saturated in both broadband $K_S$ images, so we instead analyzed the narrowband $H_2$ images.

\begin{figure*}
\begin{centering}
\epsscale{1.2}
\plotone{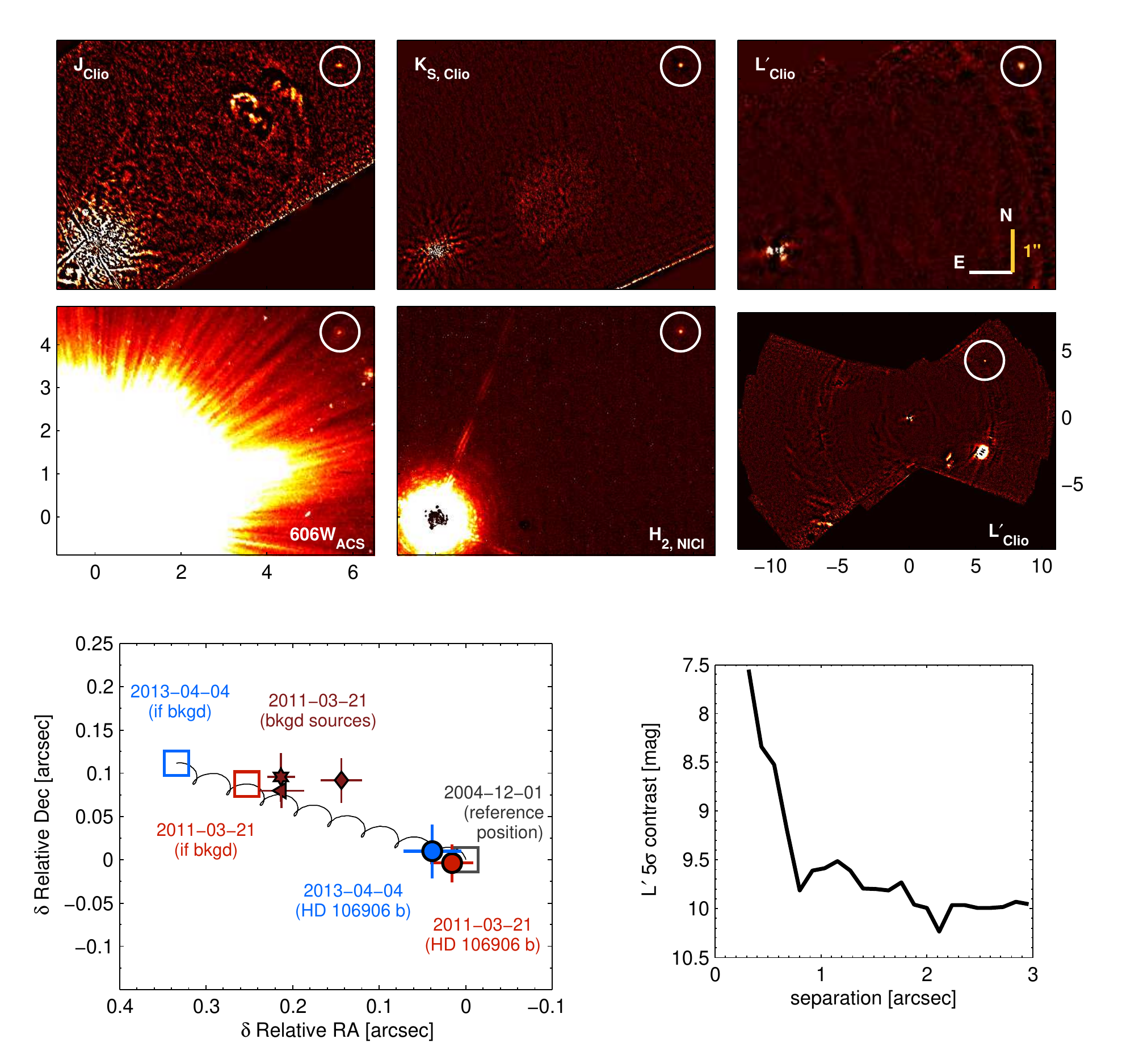}
\caption{\textit{Top row:} Clio2 $J$, $K_s$, and $L^\prime$ images, with companion circled. 
\textit{Middle row:} ACS $606W$, NICI $H_2$, and Clio2 $\Lp$ full FOV. All image scales are in arcsec as denoted in the ACS image, with the exception of the Clio2 full FOV. For visualization purposes, PSF subtraction and an unsharp mask were applied to each Clio image. 
\textit{Bottom left:} Motion in R.A. and decl. relative to HD~106906~A (a comoving object stays fixed). Coordinates of each object are normalized such that the origin (open gray square) corresponds to the object's 2004 position. Open red and blue squares denote the expected motion of a background object, filled red and blue circles the observed motion of ``b,'' and filled brown points the observed motion of three background point sources detected by both ACS and NICI.  
\textit{Bottom right:} $\Lp$ $5\sigma$ contrast curve derived from PCA analysis.  }
\label{fig:Imgs}
\end{centering}
\end{figure*}

\section{Analysis and Results}
\label{sect:A&R}

\subsection{Photometry and Astrometry}
\label{sect:PhotAst}

We determine photometry for each dataset and sensitivity for the Clio2 $\Lp$ data. We find contrasts between primary and companion of $\Delta \Lp=7.94\pm0.05$, $\Delta K_S=8.78\pm0.06$, and $\Delta J=10.7\pm0.3$ from Clio2 and $\Delta H_2=8.70\pm0.05$ from NICI; the errors are dominated by sky/background noise. For the primary's photometry, we use Two Micron All Sky Survey (2MASS) values at $J$, $H$, and $K_S$, presume $H_{2} \approx K_{S}$, and interpolate between the 2MASS and \textit{WISE} values at $\Lp$.\footnote{The primary's $K_S$, $W1$, and $W2$ magnitudes are equal within errors; we assumed the same value at $\Lp$. The magnitude derived from our standard star, HD~106965 \citep{Leggett2003}, was consistent within errors.}  Finally, we measure $[606W]=24.27\pm0.03$~mag for the companion. All photometry is listed in Table \ref{tab:props}. From our PCA reduction, we find $L^\prime$ $5\sigma$ contrasts up to $\sim10$~mag (Figure \ref{fig:Imgs}).

Three-epoch astrometry indicates that the companion is comoving. With Clio2, we find a projected separation ($\rho$) between the centroids of the primary and companion of $7\farcs11\pm0\farcs03$ at a position angle ($\theta$) of $307\fdg3\pm0\fdg2$.  With NICI and ACS, we measure $\rho=7\farcs12\pm0\farcs02$ and $\theta=307\fdg1\pm0\fdg1$, and $\rho=7\farcs135\pm0\farcs02$ and $\theta=307\fdg05\pm0\fdg1$, respectively. The expected orbital motion for a circular, face-on  $a=650$~AU orbit is $0\fdg18$, below our Clio2 astrometric precision.

The companion's astrometry is inconsistent with the expected (and observed) motion of background objects at $>6\sigma$. The proper motion of HD~106906 is $-38.79\pm0.58$~mas~yr$^{-1}$ and $-12.21\pm0.56$~mas~yr$^{-1}$ in R.A. and decl., respectively \citep{vanLeeuwen2007}. Figure \ref{fig:Imgs} shows the expected relative motion between the primary and a background object, along with astrometry for the companion and three background sources detected by both NICI and ACS (but not by Clio2).

\subsection{Companion Properties}

\begin{figure*}
\begin{centering}
\plottwo{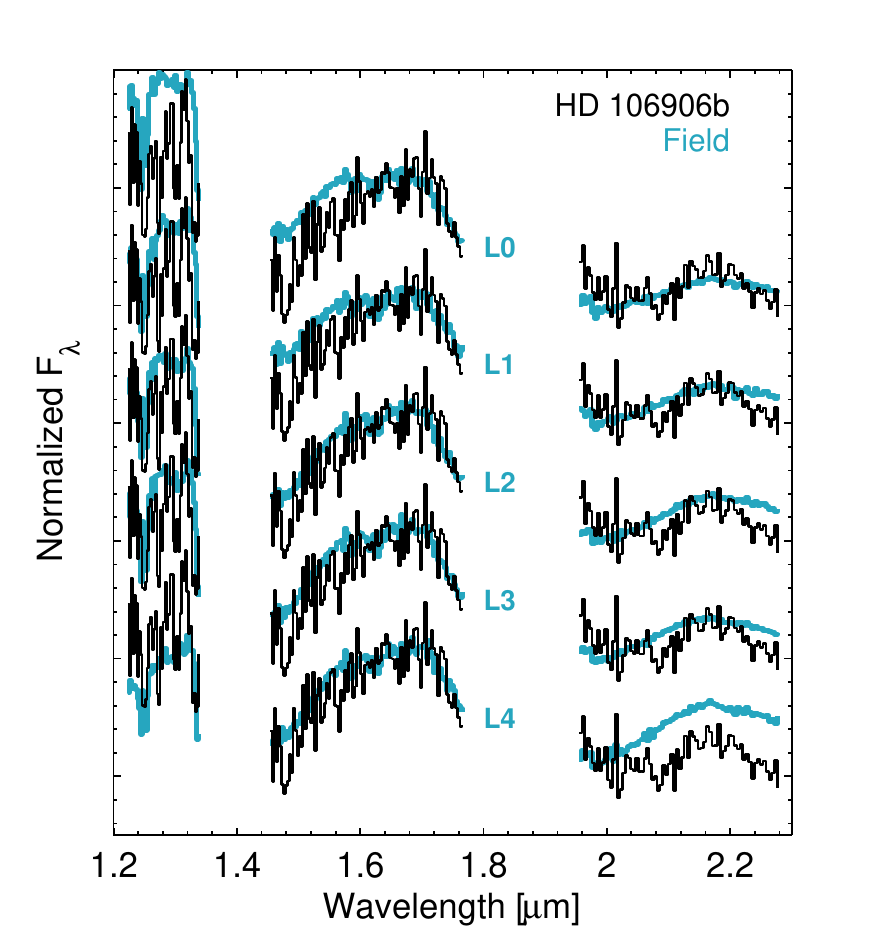}{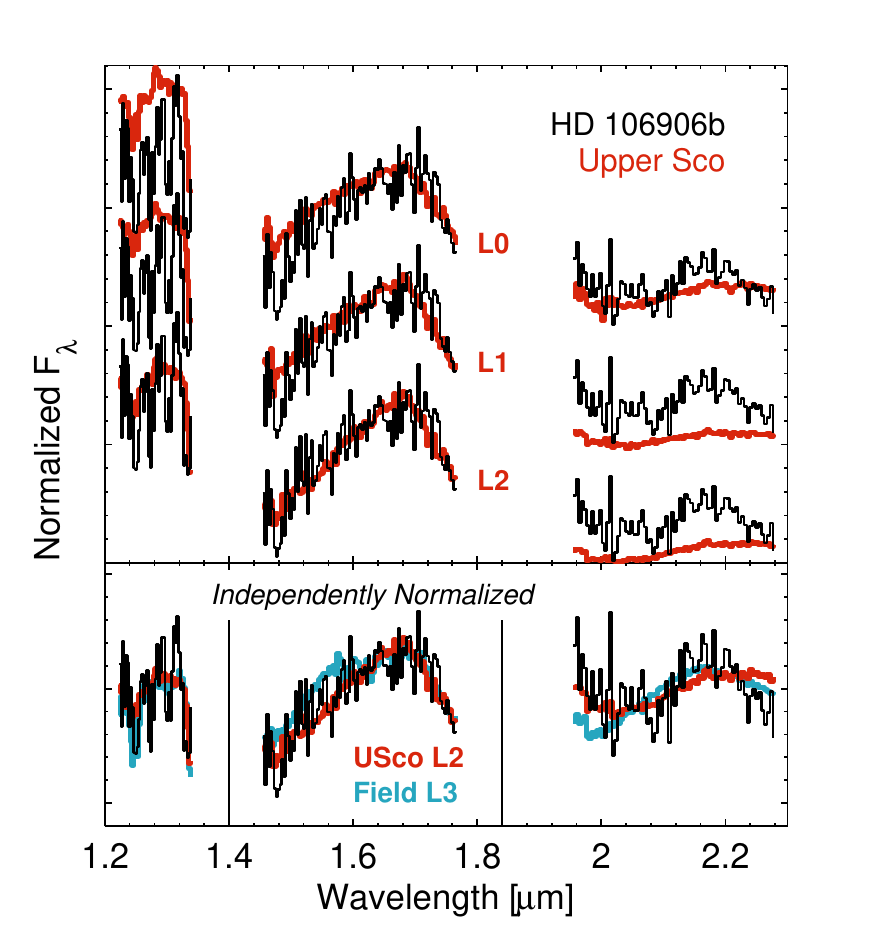}
\caption{HD~106906~b NIR SED (black line) compared to old and young BD standards, normalized at $H$-band unless otherwise noted. \textit{Left:} Field L0-L4 BDs. \textit{Right, top:} USco L0-L2 BDs. \textit{Right, bottom:} Comparison of best-fit young and field templates, with each band independently normalized. The $H$-band spectrum is best matched by the triangular shape of a low surface-gravity L2-type, although its $K$-band spectrum is better fit by a field L3-type. We tentatively classify HD~106906~b as an intermediate-gravity L$2.5\pm1$. The field objects are: 2MASSJ0746+2000AB, 2MASSJ0208+2542, Kelu-1AB, 2MASSJ1146+2230AB, and 2MASSJ2224-0158 \citep{Cushing2005}. The USco objects are: J160606-233513, J160723-221102, and J160603-221930 \citep{Lodieu2008}.}
\label{fig:spectrum}
\end{centering}
\end{figure*}

We constrain the companion's spectral type, effective temperature, luminosity, and mass based on its NIR SED and its position in color-magnitude diagrams (CMDs). Figure \ref{fig:spectrum} shows the companion's NIR SED compared to both field brown dwarfs (BDs) and young Upper Scorpius (USco) BDs. All spectra are binned to a resolving power of $\sim500$, and normalized by the average $H$-band flux (unless otherwise noted). The companion's $H$-band spectrum is somewhat triangular, and indeed it is best matched by a young L2-type dwarf. Its $K$-band spectrum, however, is rounded, more akin to that of an older field L3-type dwarf. We therefore tentatively classify HD~106906~b as an intermediate surface-gravity L2.5$\pm$1. However, a higher S/N spectrum (including the gravity-sensitive alkali lines in $J$-band) must be obtained to confirm this classification. 

We estimate the companion's mass by using a $K$-band bolometric correction (BC$_K$) to derive its luminosity, which may be compared to that predicted by the ``hot start'' COND (cloud-free) and DUSTY (cloudy) evolutionary models \citep{Baraffe2003, Chabrier2000}. We do not consider ``cold start'' models, because formation by core accretion is impossible at hundreds of AU, and scattering into the current orbit is disfavored (Section \ref{sect:interaction}). A field L2.5$\pm$1 object has $T_{eff}=1950\pm200$~K and BC$_K=3.32\pm0.13$ \citep{Golimowski2004}. Applying the field BC$_K$ yields log$(L/L_\odot)=-3.64\pm0.08$. From the evolutionary models, this corresponds to a mass of $11\pm2~M_{Jup}$ and $T_{eff}$ of $1800\pm100$~K, using an age of $13\pm2$~Myr. Even if the system is the mean age of LCC, 17~Myr, the companion remains $13~M_{Jup}$.

\begin{figure*}
\begin{centering}
\plottwo{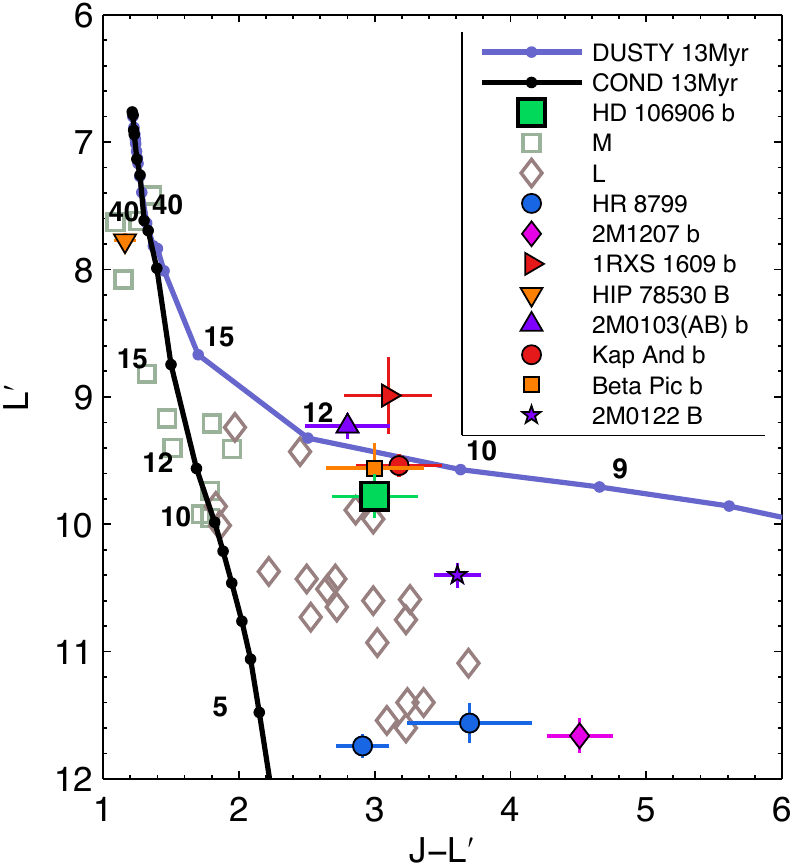}{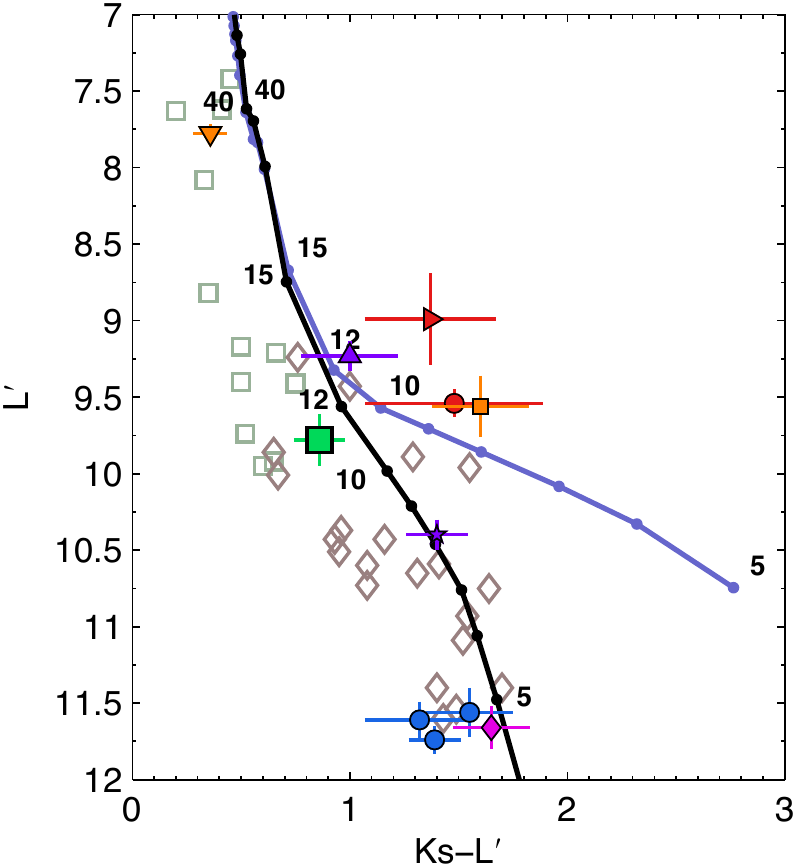}
\caption{Color magnitude diagrams plotting HD~106906~b, field M and L dwarfs \citep{Leggett2010}, and other young companions. Also plotted are DUSTY and COND models tracks for 13~Myr, with each point along the tracks corresponding to a particular mass in Jupiter masses (black text labels). The companion falls near other low-mass companions in $J/\Lp$ space, but is blue at $K_S$. The DI companions plotted are: 
HR~8799~bcde \citep{Marois2008, Marois2010}; 
2MASS~1207334-393254~b \citep[2M1207~b;][]{Mohanty2007}; 
1RXS~J160929.1-210524~b \citep[1RXS~1609b;][]{Lafreniere2008, Ireland2011, Bailey2013}; 
HIP~78530~B \citep{Lafreniere2011, Bailey2013}; 
2MASS~J01033563-5515561~(AB)b, \citep[2M0103~(AB)b;][]{Delorme2013}; 
$\kappa$~And~B \citep{Carson2013}; 
$\beta$~Pic~b \citep{Lagrange2010, Bonnefoy2011, Bonnefoy2013}; 
and 2MASS~J01225093-2439505~B \citep[2M0122~B;][]{Bowler2013}}
\label{fig:CMD}
\end{centering}
\end{figure*}

We also investigate the companion's properties using CMDs at $J$, $K_S$, and $\Lp$ (Figure \ref{fig:CMD}). For context, we plot the DUSTY and COND evolutionary model tracks as well as the photometry of field M- and L-type dwarfs \citep{Leggett2010} and other DI low-mass companions. Note that the \citet{Leggett2010} photometry are $K$-band, which is typically $\sim0.1$~mag brighter than $K_S$ for low-mass objects. In $J$ versus $L^\prime$, the companion falls on the DUSTY track, near several other low-mass companions ($\beta$~Pic~b, $\kappa$~And~B, 2M0103~B, and 1RXS~1609~b). However, it is much brighter than these objects at $K_S$, falling blue-ward of the COND track, more similar to the $K-L^\prime$ color of early- to mid-L field dwarfs. This behavior may echo that of the HR~8799 planets and 2M1207~b, which also become blue at $K_S$.

\begin{deluxetable}{lll}
\tablecaption{System properties. \label{tab:props}}
\tablewidth{0pt}
\tablehead{
\colhead{Property}	& \colhead{HD~106906~A} 	& \colhead{HD~106906~b}  
}
\startdata
Distance $(pc)$ \tablenotemark{a}   & \multicolumn{2}{c}{92$\pm$6}    \\
Age $(Myr)$ \tablenotemark{b}       & \multicolumn{2}{c}{13$\pm$2}    \\
$A_V$ \tablenotemark{b}             & \multicolumn{2}{c}{0.04$\pm$0.02} \\
$T_{eff}$          & $6516\pm165$~K \tablenotemark{b}   & $1950\pm200$~K \tablenotemark{c} \\
                  &                                    & $1800\pm100$~K \tablenotemark{d} \\
Spectral type     & F5V \tablenotemark{b}              & L2.5$\pm$1 \\
log($L/L_\odot$)   & 0.75$\pm$0.06 \tablenotemark{b}    & $-3.64\pm$0.08 \\
Mass           	 & $1.5\pm0.10~\msun$\tablenotemark{b} & $11\pm2~M_{Jup}$ \\
Separation $('')$ &   \multicolumn{2}{c}{7.11$\pm$0.03}    \\
PA $(\degr)$      &   \multicolumn{2}{c}{307.3$\pm$0.2}	   \\
$606W$ 	          & $-$	                                & 24.27$\pm$0.03 \\
$J$ 	          & 6.95$\pm$0.03 \tablenotemark{e}	    & 17.6$\pm$0.3  \\
$K_S$ 	          & 6.68$\pm$0.03 \tablenotemark{e}    & 15.46$\pm$0.06  \\
$H_2$             & 6.68$\pm$0.05 \tablenotemark{f}     & 15.38$\pm$0.07   \\
$\Lp$             & 6.7$\pm$0.1            & 14.6$\pm$0.1     \\
$W1$ \tablenotemark{g}  & \multicolumn{2}{c}{6.68$\pm$0.04}   \\
$W2$ \tablenotemark{g}  & \multicolumn{2}{c}{6.68$\pm$0.02}   \\
$W3$ \tablenotemark{g}  & \multicolumn{2}{c}{6.59$\pm$0.02}   \\
$W4$ \tablenotemark{g}  & \multicolumn{2}{c}{4.66$\pm$0.03}   \\
\enddata

\tablenotetext{a}{\textit{Hipparcos} catalog \citep{vanLeeuwen2007}.}
\tablenotetext{b}{\citet{Pecaut2012}}
\tablenotetext{c}{Effective temperature from field dwarf scale.}
\tablenotetext{d}{Effective temperature from evolutionary models.}
\tablenotetext{e}{2MASS $J/H/K_S$ survey. Unresolved.}
\tablenotetext{f}{Assumed equal to 2MASS $K_S$.}
\tablenotetext{g}{\textit{WISE} survey. Unresolved.}

\end{deluxetable}

\subsection{Bound Companion or Free-Floating Cluster Member?}

We calculate the probability that HD~106906~b is not bound, but instead a free-floating cluster member with similar proper motion, by estimating the space density of free-floating BDs in LCC. Because the census of BD cluster members is incomplete, we extrapolate their space density from the known B star population. We take the 44 known B star cluster members from the \textit{Hipparcos} catalog \citep{deZeeuw1999}, plus an additional 11 likely members, which we believe were spuriously rejected from the catalog because their space motions are perturbed by binary companions. Assuming the census of B stars is complete, a Kroupa initial mass function \citep{Kroupa2001} predicts a total stellar population of 1836 stars above $0.08~\msun$. 

We next scale the total stellar population by an assumed BD fraction. Surveys of young clusters have found $N_{BD}/N_{star}\sim0.2$ \citep[e.g.:][]{Slesnick2004, Luhman2007}. If LCC has a similar ratio, then it should contain $\sim370$ BDs below $0.08\msun$. Most cluster members are concentrated within $\sim500$ deg$^2$ \citep{deZeeuw1999}.  HD~106906 is one of two LCC members observed in our disk-selected program\footnote{Interestingly, the other system, HD~95086, also hosts a DI planetary-mass companion \citep{Rameau2013a}.}, hence the probability of a chance alignment within 7'' is $<1\times10^{-5}$. We conclude that HD~106906~Ab is most likely a bound pair.

\subsection{Constraints on Additional Companions}
\label{sect:additionalobj}

No additional point sources are detected in our $\Lp$ image. We could detect additional companions as massive as ``b'' at projected separations $>0\farcs38$, reaching a background limit as low as $4~M_{Jup}$ (based on COND models). At larger separations, we achieve a sensitivity of $5-7~M_{Jup}$.  Two low S/N NICI sources ($\rho=9\farcs6$, $\theta=236\degr$ and $\rho=7\farcs1$, $\theta=76\degr$) do not have counterparts in the \textit{HST} or Clio2 images (the first is not within the Clio2 $\Lp$ FOV). The sources have $K_S=19.5-20$, based on their contrast with HD~106906~b. From the available data, we cannot determine the nature of either faint NICI source.

\subsection{Circumstellar Disk Properties and Companion-Disk Interaction}
\label{sect:interaction}

HD~106906 was selected for DI because it has a large IR excess indicative of a massive debris disk, and because the shape of the excess' SED suggests the disk is devoid of both hot and warm material. \citet{Chen2005} derived a color temperature of 90~K and $L_{IR}/L_\star=1.4\cdot10^{-3}$ from 24 and 70$\um$ broadband photometry. With the addition of {\it Spitzer} Infrared Spectrograph and MIPS-SED spectra, we confirm that the disk emission is well fit by a blackbody temperature of $95$~K, using data up to $\sim100\um$.

As the disk is not resolved at any wavelength, we estimate its plausible extent from the dust temperature. Given the stellar parameters of HD~106906~A (Table \ref{tab:props}), the inferred dust location is $\sim$20 AU, assuming blackbody-like grains. The inner radius ($r_{in}$) could be as close as 15~AU for silicate grains with radii of 10$\um$. Using a sample of nine \textit{Herschel} resolved disk images around early-type stars, \citet{Booth2013} showed that the measured disk sizes are $1-2.5$ times larger than the blackbody estimates. The discrepancy increases toward later spectral types, reaching a factor of $\sim$6 for a G5V-type host \citep{Wyatt2012}. Given the primary's F5V spectral type, the outer radius of the disk ($r_{out}$) is likely to be $<$120~AU. We adopt a model with a $20-120$~AU dust ring for the following discussion. Future resolved imaging is required to determine the true extent of the disk.

The HD~106906 system adds to a small but growing sample of DI planetary systems with debris disks. It has been suggested that the planets in these systems play a hand in sculpting their debris disks \citep[e.g.: HR~8799, $\beta$~Pic, HD~95086:][]{Su2009, Lagrange2010, Moor2013}. HD~106906~b could similarly be shepherding the disk around its primary star if it is on an eccentric orbit; a massive companion will disrupt disk material between its Hill sphere at periastron and its Hill sphere at apastron. To gravitationally sculpt the disk's outer edge at 120~AU, the companion's periastron must be 135~AU. Presuming it is at apastron now, the orbit would require an eccentricity of 0.65. If the disk's outer radius is smaller, the companion is not currently at apastron, or the orbit is inclined relative to our line of sight, the necessary orbital eccentricity would increase. 

Two formation mechanisms are typically postulated for wide planetary-mass companions: in situ formation (binary-star-like) or formation in a tight orbit followed by scattering to a wide orbit (planet-like). Scattering from a formation location within the current disk is unlikely to have occurred without disrupting the disk in the process \citep{Raymond2012}. We also note that the perturber must be $>11~M_{Jup}$; we do not detect any such object beyond 35~AU (Section \ref{sect:additionalobj}), disfavoring formation just outside the disk's current outer edge. While it is possible that the companion is in the process of being ejected on an inclined trajectory from a tight initial orbit, this would require us to observe the system at a special time, which is unlikely. Thus we believe the companion is more likely to have formed in situ in a binary-star-like manner, possibly on an eccentric orbit \citep{Duquennoy1991}. However, binary mass ratios of $M_B/M_A<10\%$ are rare \citep{Reggiani2013}, and in this system $M_b/M_A<1\%$, so the formation process remains somewhat ambiguous.

\section{Summary}
\label{sect:summary}
At $11\pm2~M_{Jup}$, HD~106906~b is the first planetary-mass companion discovered with the new MagAO/Clio2 system, underscoring the power of $\Lp$ surveys for detecting low-mass companions and discriminating against background contaminants. We have confirmed its cool (spectral type L$2.5\pm1$), young nature with NIR spectroscopy and its comoving status with astrometry over an 8.3~yr baseline. At 650~AU projected separation, this is one of the widest planetary-mass companions known.

HD~106906~A was targeted because it hosts a massive, ring-like debris disk, potentially sculpted by a planetary companion. From the disk's SED, we estimate $r_{in}\sim20$~AU and $r_{out}\lesssim120$~AU. The companion would require an orbital eccentricity $>0.6$ to gravitationally sculpt the outer edge of this disk. The presence of a massive disk around the primary argues against a scattering origin for the companion. We suggest it is more likely to have formed in situ in a binary-star-like process, though $M_b/M_A<1\%$ is unusually small.

HD~106906~b joins a growing sample of widely separated, planetary-mass and BD companions whose formation mechanisms are poorly understood. Each additional example is valuable, particularly when additional environmental information is present (such as the existence and morphology of a circumstellar disk). Future scattered light or sub-mm observations of the HD~106906 circumprimary disk might uncover signs of dynamical instabilities and further constrain the system's formation process.

\acknowledgments

We are grateful to the MagAO development team and the Magellan Observatory staff for their support. We thank Rob Simcoe for FIRE pipeline assistance and the referee for useful feedback.
This Letter includes data from the Magellan Telescopes, archival observations from the Hubble Space Telescope and Gemini Observatory, and photometry from the 2MASS and WISE surveys.
 MagAO was funded by NSF MRI, TSIP, and ATI awards. VB acknowledges the NSF Graduate Research Fellowship Program (DGE-1143953). EEM acknowledges NSF grants AST1008908 and AST1313029. KM and JM are partially supported by the NASA Sagan Fellowship Program via Caltech. AJW acknowledges funding from NAI through Cooperative Agreement NNA09DA81A

\end{document}